\documentclass[11pt, final, 3p, times]{elsarticle}

\usepackage{lipsum}

\usepackage{bm}
\usepackage{booktabs}
\usepackage{multirow}
\usepackage{array}
\usepackage{listings}
\usepackage{hyperref}
\usepackage{amsmath}
\usepackage[dvipsnames]{xcolor}

\usepackage{float}
\usepackage[export]{adjustbox}

\usepackage[figuresright]{rotating}
\makeatletter
\newcommand\extralabel[2]{{\edef\@currentlabel{\@currentlabel#2}\label{#1}}}
\makeatother

\newcolumntype{M}[1]{>{\centering\arraybackslash}m{#1}}
\usepackage{url}

\usepackage{capt-of}
\usepackage{booktabs}
\usepackage{varwidth}

\usepackage{etoolbox}
\makeatletter
\patchcmd{\ps@pprintTitle}
  {Preprint submitted to}
  {Accepted - Biomedical Signal Processing and Control -}
  {}{}
\makeatother

\begin{document}

\begin{frontmatter}

\title{Robust Deep Learning for Eye Fundus Images: Bridging Real and Synthetic Data for Enhancing Generalization}
    
\author[1,2]{ Guilherme C. Oliveira}
\ead{gc.oliveira@unesp.br}

\author[1]{ Gustavo H. Rosa}
\ead{gustavo.rosa@unesp.br}

\author[1]{ Daniel C. G. Pedronette}
\ead{daniel.pedronette@unesp.br}

\author[1]{ Jo\~{a}o P. Papa}
\ead{joao.papa@unesp.br}

\author[3]{ Himeesh Kumar}
\ead{himeesh@gmail.com}

\author[1]{ Leandro A. Passos}
\ead{leandro.passos@unesp.br}

\author[2]{ Dinesh Kumar}
\ead{dinesh@rmit.edu.au}

\address[1]{{School of Sciences, S\~ao Paulo State University},
    {S\~ao Paulo},
    {Brazil}}
    
\address[2]{{School of Engineering, Royal Melbourne Institute of Technology},
    {Victoria},
    {Australia}}
    
\address[3]{{Centre of Eye Research, University of Melbourne},
    {Victoria},
    {Australia}}

\begin{abstract}
Deep learning applications for assessing medical images are limited because the datasets are often small and imbalanced. The use of synthetic data has been proposed in the literature, but neither a robust comparison of the different methods nor generalisability has been reported. Our approach integrates a retinal image quality assessment model and StyleGAN2 architecture to enhance Age-related Macular Degeneration (AMD) detection capabilities and improve generalizability.
This work compares ten different Generative Adversarial Network (GAN) architectures to generate synthetic eye-fundus images with and without AMD.  We combined subsets of three public databases (iChallenge-AMD, ODIR-2019, and RIADD) to form a single training and test set. We employed the STARE dataset for external validation, ensuring a comprehensive assessment of the proposed approach.
The results show that StyleGAN2 reached the lowest Fréchet Inception Distance (166.17), and clinicians could not accurately differentiate between real and synthetic images. ResNet-18 architecture obtained the best performance with 85\% accuracy and outperformed the two human experts (80\% and 75\%) in detecting AMD fundus images. The accuracy rates were 82.8\% for the test set and 81.3\% for the STARE dataset, demonstrating the model's generalizability. The proposed methodology for synthetic medical image generation has been validated for robustness and accuracy, with free access to its code for further research and development in this field.

\end{abstract}


\end{frontmatter}

\section{Introduction}

Deep learning (DL) techniques are suitable for automated analysis of medical images and have been shown to outperform experts in several fields. Examples of applications of such approaches are to perform discriminative retinal image analysis, such as automated classification of fundus images for detection of referable diabetic retinopathy~\cite{gulshan2016development,quellec2017deep,gargeya2017automated,de2018clinically, khojasteh2018introducing}, retinopathy of prematurity~\cite{brown2018automated}, exudates on the retina~\cite{khojastehCBM:2019}, and age-related macular degeneration (AMD)~\cite{burlina2019assessment,burlina2017automated,burlina2018utility}, and also for granular AMD severity classification~\cite{burlina2018use,grassmann2018deep}. However the potential of this approach for medical images is limited because they require a large number of annotated images that are not currently available. It is also essential that the datasets should be balanced to avoid biased training. Usually, annotated medical image data sets are not sufficiently large nor balanced, and this is often due to issues of privacy of medical data~\cite{esteva2019guide}.

\begin{figure}[!ht]
    \centering
    \includegraphics[scale=0.11]{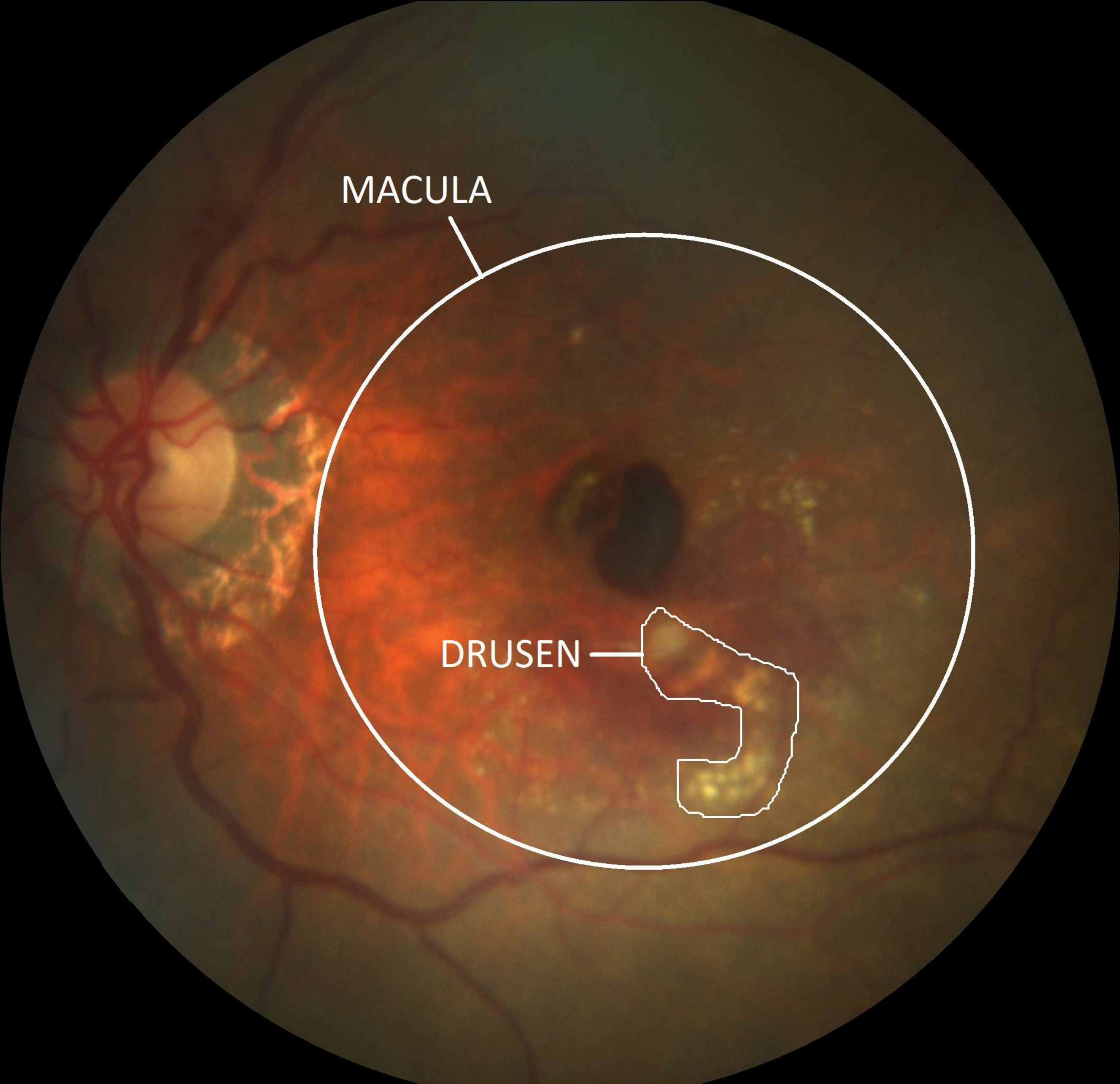}
    \caption{ Retina fundus image positive to age-related macular degeneration identified by the presence of drusen. The image was extracted from the iChallenge-AMD dataset~\cite{fu2020adam}.}
    \label{f.amd_example}
\end{figure}

Different approaches have been used to address this shortcoming such as transfer learning using previously trained models, and data augmentation~\cite{khojasteh2018fundus, khojasteh2019novel}. Brigato et al.~\cite{brigato2021close} highlight the value of standard data augmentation techniques, such as cropping, rotation, and color jittering, as well as using low-complexity convolutional neural networks for small datasets. However, they suggest the development of more complex data generation and augmentation pipelines. Spatial transformations, while adding images, do not guarantee data variability~\cite{perez2017effectiveness}, and pre-trained networks often face limitations due to being trained in different learning domains.

\begin{sloppypar}
A promising solution to generate synthetic images lies in the Generative Adversarial Networks (GANs)~\cite{goodfellow2014generative}. Such an approach performs data augmentation by competitively creating new samples, i.e., a generator attempts to create synthetic images to fool the discriminator, which then tries to identify whether they are fake or real. GANs have also provided exceptional results over a wide variety of medical applications, such as liver lesion classification~\cite{frid2018gan}, Barrett's esophagus identification~\cite{Souza2020:CBM,souza2021fine}, and chest pathology recognition~\cite{salehinejad2018generalization}. However, these did not address problems with generalization, where there is a variable quality and other differences in imgaes of the dataset. 
\end{sloppypar}

In the field of Optical Coherence Tomography (OCT) imaging, super-resolution GANs (like ESRGAN~\cite{wang2018esrgan}) have demonstrated their value as a tool to enhance the quality of the image and improve AMD detection~\cite{thakoor2022enhancing}. Das et al.~\cite{das2020unsupervised} proposed a quick and reliable super-resolution approach concerning OCT imagery using GANs, achieving a classification accuracy of $96.54\%$.

In retinal imaging, GANs have been used to create synthetic data. Li et al.~\cite{li2021applications} highlighted the importance of enhancing the quality of synthetic retinal images in their review, emphasizing that using synthetic images in training can improve performance and help mitigate overfitting.
Bellemo et al.~\cite{bellemo2018generative} described the possible advantages and limitations towards synthetic retina image generation using GANs. The authors highlighted the potential clinical applications of GANs concerning early- and late-stage AMD classification. 
Burlina et al.~\cite{burlina2019assessment} trained a Progressive GAN~\cite{karras2017progressive} on $133,821$ color fundus images from $4,613$ age-related eye disease individuals to learn how to generate synthetic fundus images with and without AMD. Two retina specialists were asked to distinguish between images with and without AMD for original and synthetic images. Recognition rates varied from around $84\%$ for the first specialist to about $89\%$ for the second. The accuracy differences between synthetic and real images did vary slightly for both specialists. While the outcomes show great potential, the authors did not verify the utilization of data augmentation during the training process with their approach. Furthermore, Burlina et al.~\cite{burlina2019assessment} conducted their research using the Age-Related Eye Disease Study (AREDS) dataset. This dataset mainly includes participants from the United States, making it reflective of a North American demographic.

Anh et al~\cite{ahn2023fundusgan} tested the FundusGAN to generate eye-fundus images for two eye disease: Age-related macular degeneration and Diabetic retinopathy and demonstrated its ability for the synthetic images to be generalisable for the two disease.However, Anh et al.~\cite{ahn2023fundusgan} work was confined to a single dataset in which most participants were from India and lacked diversity. Thus, they could not evaluate their technique across different ethnicities, demographics, and equipment variations, and therefore, their study could not be tested for generalisability.

StyleGAN2 has been utilized to generate detailed images of the eye fundus. Mayya et al.~\cite{mayya2023empirical} performed data augmentation with StyleGAN2. They developed a method to diagnose a range of conditions such as myopia, diabetic retinopathy, age-related macular degeneration, glaucoma, and cataract. Meanwhile, Wang et al.~\cite{wang2023synthetic} employed a distinct approach using StyleGAN2 and augmentation, categorizing conditions into four classes: no AMD, early AMD, intermediate AMD, and advanced AMD. However, Mayya et al.~\cite{mayya2023empirical} and Wang et al.~\cite{wang2023synthetic} didn't specify a method for removing images with significant quality issues, such as those that are blurry, have low contrast, or are inadequately illuminated. Veturi et al.~\cite{veturi2023syntheye} conducted a diagnostic study focusing on gene-labeled fundus autofluorescence (FAF) images for inherited retinal diseases (IRDs). However, They found that synthetic data augmentation failed to enhance disease classification when applied to IRD datasets.

Age-related macular degeneration (AMD) is a major cause of vision impairment and has affected approximately $200$ million people worldwide in 2020~\cite{jonas2014global}. With an aging population, such numbers are expected to rise to $288$ million by 2040~\cite{wong2014global}. AMD is a progressive disorder of the macular region that causes central vision loss and is one of the most common causes of irreversible vision impairment in people over $50$ years-old~\cite{harvey2003common}.  Figure~\ref{f.amd_example} depicts an example of a retina fundus affected by AMD.

This work has introduced an alternative approach for generating synthetic images for training deep networks and tested it for AMD identification, which consists in using a retinal image quality assessment model~\cite{fu2019evaluation} and the StyleGAN2-ADA~\cite{Karras2020ada}. Retina images, positive and negative to AMD, from multiple databases having a range of image qualities and lesions were used. Ten different GAN architectures were compared to generate synthetic eye-fundus images and the quality was assessed using the Fréchet Inception Distance (FID), two independent clinical experts who were label blinded and deep-learning classification. Different percentages of synthetic data were employed in the augmentation.

The primary contributions of this work are fourfold:

\begin{itemize}
\item To introduce StyleGAN2-ADA~\cite{Karras2020ada} for eye fundus image generation; 
\item Test generalization across different datasets;
\item Free access to software for generating the synthetic images;
\item Accessible web-based tool for diagnosing AMD that combines computer vision and deep learning techniques.
\end{itemize}

The remainder of this paper is organized as follows. Section~\ref{s.methodology} presents  methodology. Section~\ref{s.experimental_results} describes the experimental results and Section~\ref{s.conclusion} states conclusions and future works.

\section{Methodology}
\label{s.methodology}

This section describes the datasets that were used in the study, the techniques employed to generate the synthetic images, and the methodology to evaluate the different neural architectures considered in the experimental section.  

\subsection{Dataset}
\label{ss.datasets}

This study used eye-fundus images from four public datasets from four countries which have been reported in the literature. These capture the typical differences due to equipment and demographics, which is necessary for verifying the generalisability of the data. Below, we summarized the datasets' preliminary information:

\begin{itemize}
    \item \textbf{iChallenge-AMD:} Comprises of $1,200$ retinal fundus images that have been annotated for drusen and hemorrhage. The training set was made of 400 images (89 images of eyes with AMD and 311 from eyes without AMD), while the test sets contained the remaining images~\cite{fu2020adam}. 
    
    \item \textbf{ODIR-2019:} Contains colored fundus images from both left and right eyes of $5,000$ patients obtained from multiple hospitals/medical centers in China, with varying image resolutions and observations from several specialists. The dataset has was designed to address normal and six diseases: diabetes, glaucoma, cataract, AMD, hypertension, myopia, and other diseases/abnormalities\footnote{\url{https://odir2019.grand-challenge.org/dataset/}}. The training set is made up of a structured dataset with $7,000$ images, of which $280$ images are labelled as having AMD. The testing set consists of $500$ colored fundus images, eliminating age and gender. 
    
    \item \textbf{RIADD:} contains $3,200$ fundus images recorded using 3 different cameras and  multiple conditions. The images have been annotated through the consensus of two retina experts. The dataset has been sub-divided based on six diseases/abnormalities; diabetic retinopathy, AMD, media haze, drusen, myopia, and branch retinal vein occlusion~\cite{pachade2021retinal}. The dataset was subdivided into three subsets: $60\%$ for training ($1,920$ images), $20\%$ for validation ($640$ images), and the remaining $20\%$ for testing purposes ($640$ images). 

    \item \textbf{STARE:}  STructured Analysis of the REtina (STARE), contains 400  high-quality fundus images with different diseases.\footnote{\url{https://cecas.clemson.edu/~ahoover/stare/}}. We randomly sampled 40 AMD and 40 normal images from STARE for testing our model. This database was not part of the mix that was used to train the model and hence testing the model from images from this database was a test for its generalisability.  All the ground truths are annotated by the expert. 
  
    We have used publicly available and online datasets, each of which have described their patient recruitment, clinical outcomes and human experiments ethics approval in their associated publications. Registration was necessary for access to the data from the three sources.
    
    i-Challenge datset authors confirm in their publications they received ethics approval  guidelines from Sun-Yat Sen University, China and Sixth People’s Hospital Affiliated to Shanghai Jiao Tong University, Shanghai, China. Details on https://amd.grand-challenge.org/Home/  
    
    ODIR dataset confirm that all data is from routine clinical examinations, have been de-identified and published after receiving clearance from Peiking University board. Details on https://odir2019.grand-challenge.org/dataset/
    
    RIADD dataset confirms that data collection was conducted after ethics approval from Review Board of Shri Guru Gobind Singhji Institute of Engineering and Technology, Nanded, India. Details on https://riadd.grand-challenge.org/

\end{itemize}

The datasets mentioned above were designed to address a challenge and hence the labels of the test subset were not available. Therefore, we used only the training subset.
 
These datasets have an imbalanced class distribution, which is an inherent problem for most medical image datasets and using such a set generally leads to a bias towards the predominant class. Different research groups have developed these datasets with differences in the quality of the images and the demographics. Thus, these datasets offered both, data imbalance and wide range of image quality. 

\begin{figure}[!ht]
  \centering
  \begin{tabular}{ccc}
        \includegraphics[width=4.8cm,height=3.5cm]{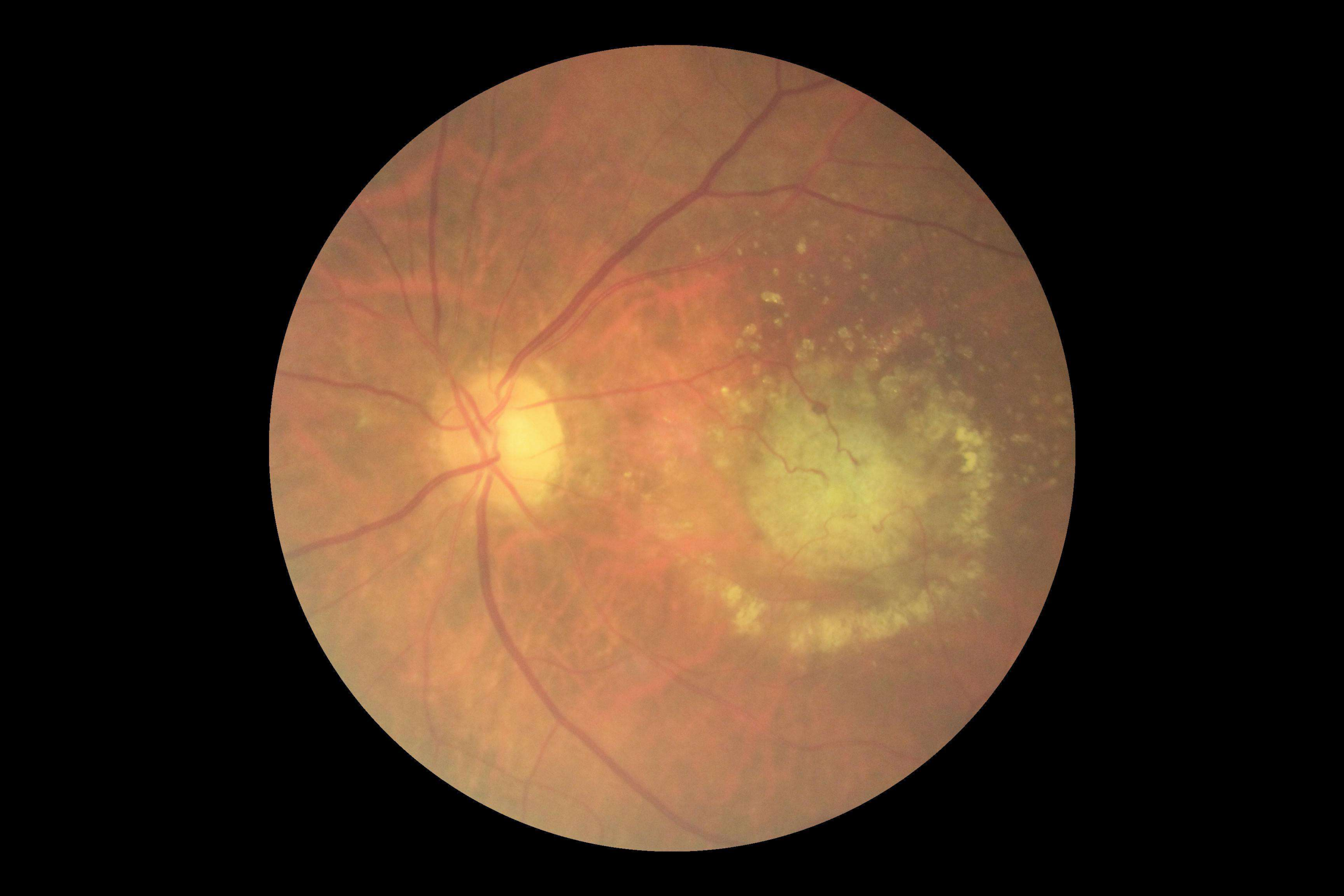} & 
        \includegraphics[width=3.5cm,height=3.5cm]{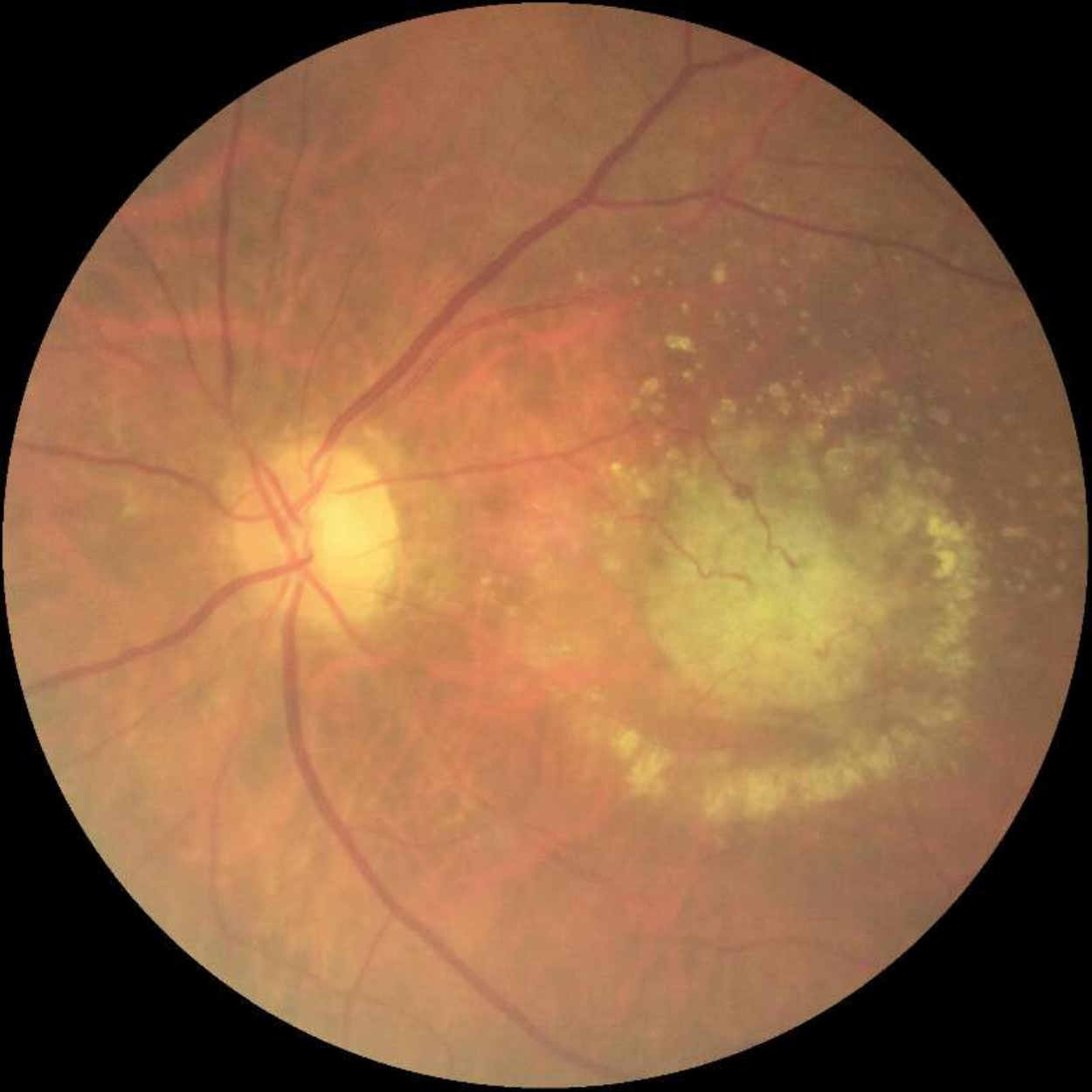} &
        \includegraphics[width=3.5cm,height=3.5cm]{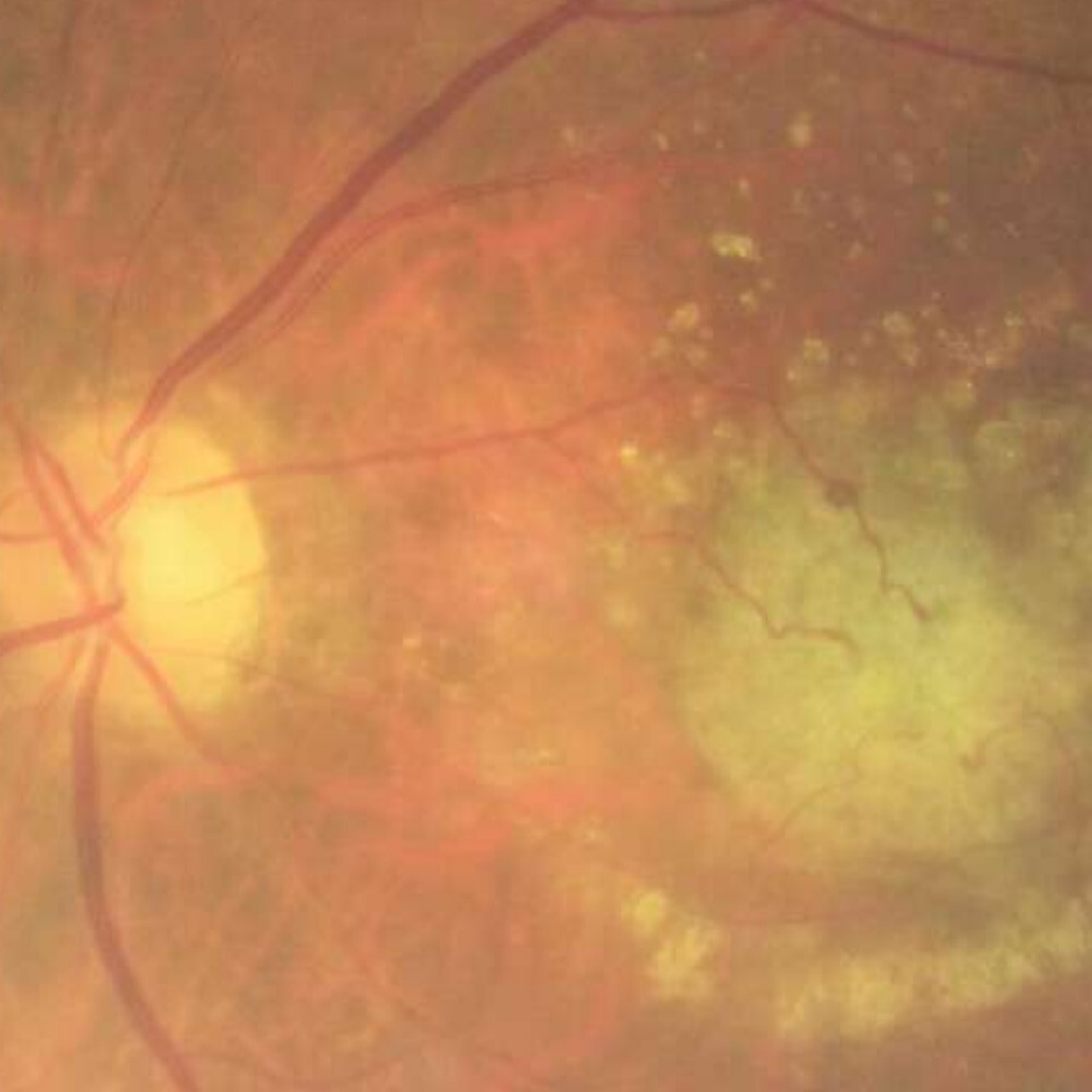} \\ 
        (a) & (b) & (c) 
    \end{tabular}
    \caption{ Sample image extracted from ODIR-2019 dataset and its corresponding transformations: (a) original image, (b) background removal using Hough Circle Transform and resizing, and (c) central cropping.}
    \label{f.preprocess}
\end{figure}

The ODIR-2019 and RIADD datasets were organized into two subsets, AMD and non-AMD images.
Preprocessing methodology and quality classification was same as proposed by Fu et al.~\cite{fu2019evaluation} , which comprises of a step that detects the retinal mask using the Hough Circle Transform and then crops it to remove the impact of the black background. The cropped region was then scaled to a $224\times224$ resolution. 
The resultant images were submitted to a DenseNet trained by Fu et al.~\cite{fu2019evaluation} which classified these in 3 classes: “Good”, “Usable” and “Reject”. The rejected images were of poor quality based on significant blurring, low contrast, inadequate illumination, which has been described by Fu et al.~\cite{fu2019evaluation}. As a result, the number of images positive to AMD decreased from $89$ to $74$, $280$ to $227$, and $100$ to $79$ images in the iChallenge-AMD, ODIR-2019, and RIADD datasets, respectively while the number of non-AMD images decreased from $311$ to $290$, $6,720$ to $4,993$, and $1,820$ to $1,143$ images concerning the iChallenge-AMD, ODIR-2019, and RIADD datasets, respectively. Figure~\ref{f.preprocess} illustrates the steps mentioned above for a sample image from the ODIR-2019 dataset.

These images were then resampled to $390\times390$ pixels, followed by a cropping procedure keeping the center of the image to $256\times256$ pixels. Such a procedure is required to drive StyleGAN2-ADA generating images focused on the macula area (Figure~\ref{f.preprocess}-c). Ultimately, images were resized  $224\times224$ pixels and normalized within the range $[-1,1]$ to be used as proper inputs to the deeper architectures considered in the manuscript.

\begin{figure}[!ht]
  \centering
  \begin{tabular}{cc}
        \includegraphics[scale=0.37]{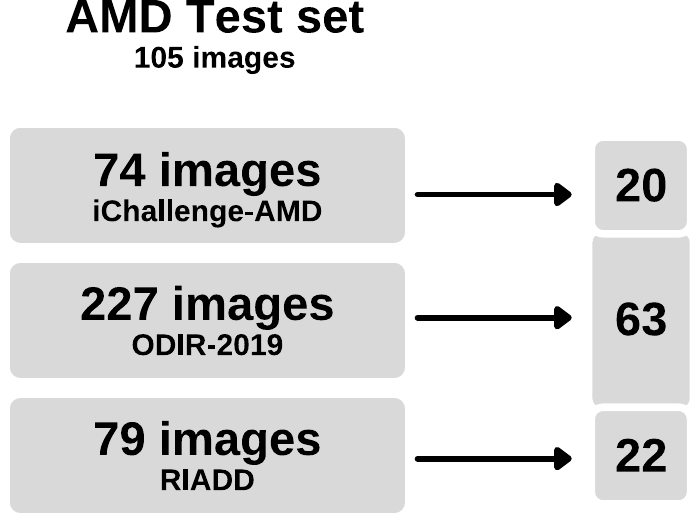} & 
        \includegraphics[scale=0.37]{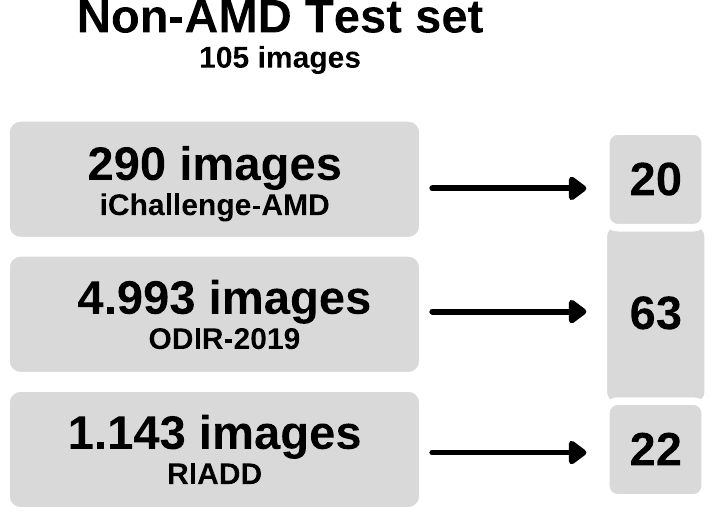} \\ 
        (a) & (b)
    \end{tabular}
\caption{ Number of images per dataset to compose the test set: (a) images positive to AMD and (b) non-AMD images.}
\label{f.val_set}
\end{figure}

After quality assessment and selecting the images for the final dataset using the criterion  described earlier, the resulting single dataset comprised totaL of $7,106$ images. In this, $6,896$ images were used to train the models ($275$ with AMD and $6,621$ without AMD) and the remaining $210$ images ($105$ with AMD) were used as a holdout test set. Figure~\ref{f.val_set} displays the number of images used per dataset to compose the final test set. 

\subsection{Evaluation Measures}
\label{ss.evaluation_metrics}

Evaluating the quality of synthetic images is important for establishing their usability in practical applications, such as training deep learning models. It can significantly influence the training of these models. If the data does not accurately represent reality or lacks diversity, the synthetic data may introduce noise into the training, decreasing the model performance.

We employed three evaluation measures, i.e., the Fréchet Inception Distance (FID), a well-known GAN evaluation score~\cite{heusel2017gans}, the ability of human experts to identify the synthetic images and the classification accuracy. FID is often used to assess the quality and variety of the generated images and, even though it has been proposed to improve the standard Inception Score, it still uses the Inception architecture to extract features from both, synthetic and real images. 

Structural Similarity Index (SSIM), and Peak Signal-to-Noise Ratio (PSNR) focus on pixel-wise comparisons and are limited in their capacity to assess higher-level features and perceptual quality. FID leverages the features extracted from a pre-trained Inception Network to measure the similarity of feature representations between real and generated images.

\subsection{Experimental Setup}
\label{ss.experimental_setup}

There were four stages of the experiments: (i) comparison of ten different GAN architectures, (ii) evaluation of synthetic images based on human experts' ability to distinguish between real and synthetic images, (iii) evaluation of data augmentation with synthetic images in three different deep learning networks, and (iv) measuring the accuracy in identifying AMD images between human experts and deep learning networks trained with mixed data, with both real and synthetic images.

\begin{sloppypar}
The first experiment employed FID to compare StyleGAN2-ADA and nine distinct GAN models. The following models were considered in the experiment: Deep Convolutional GAN (DCGAN)~\cite{radford2015unsupervised}, Least Squares Generative Adversarial Networks (LSGAN)~\cite{mao2017least}, Wasserstein GAN (WGAN)~\cite{arjovsky2017wasserstein}, Wasserstein GAN with Gradient Penalty (WGAN-GP)~\cite{gulrajani2017improved}, Deep Regret Analytic Generative Adversarial Networks (DRAGAN)~\cite{kodali2017convergence}, Energy-based Generative Adversarial Network (EBGAN)~\cite{zhao2016energy}, Boundary Equilibrium Generative Adversarial Networks (BEGAN)~\cite{berthelot2017began}, Conditional GAN (CGAN)~\cite{mirza2014conditional}, and Auxiliary Classifier GAN (ACGAN)~\cite{odena2017conditional}. All models were trained with $50$ epochs, considering samples of size $100\times100$ pixels and a batch size of $32$. The training step used the ADAM\cite{kingma2017adam} optimizer with a learning rate of $0.0002$ and decay rates of $0.5$ and $0.999$ regarding the generator and the discriminator, respectively. The experiments were conducted using the training set with an Nvidia RTX 2060 GPU. Therefore, after training each GAN model, new images were generated and FID was computed.
\end{sloppypar}

The second experiment determines whether clinical experts, who are very experienced with the analysis of eye fundus images, can distinguish between synthetic and real images. Such a step is essential to evaluate the effectiveness of StyleGAN2-ADA for generating synthetic eye fundus images. The experts were provided with randomly generated image sets, one for AMD-diagnosed images and the other for non-AMD images, each consisting of ten synthetic images and ten real images. They were asked to identify the syntetic images in the mix. 

In the next experiment, we considered three deep architectures pre-trained with ImageNet dataset~\cite{deng2009imagenet} for performance comparison when the model is augmented with synthetic images, i.e., SqueezeNet~\cite{iandola2016squeezenet}, AlexNet~\cite{krizhevsky2012imagenet}, and ResNet18~\cite{he2016deep}. During training, synthetic and real images were mixed within each batch according to a pre-determined hyperparameter $p\in[0,1]$. For each image, a uniform distributed number $x$ was sampled and compared with $p$: if the latter was greater than $x$, the image was replaced by a synthetic one. We considered images generated by StyleGAN2-ADA, for it had obtained the best FID values in the first experiment (such outcomes are later described in Section~\ref{s.experimental_results}). The deep networks were trained using a learning rate of $0.0001$, a decay rate of $0.9$, batch size of $32$, number of epochs equal to $5$, and samples of size $224\times224$ pixels.

To address the issue of unbalanced dataset, Weighted Random Sampler~\cite{Efraimidis2008} was used. This approach can be implemented in PyTorch~\cite{paszke2019pytorch}. It assigns weights to each data point in the dataset. These weights are inversely proportional to the class frequency to which they belong. When batches of data are drawn for training the model during the training process, this sampler uses these weights to influence the selection process, making it more likely for points from minority classes to be included in each batch. Further augmentation was performed using classical image transformations, such as resizing and color jittering, which changes the brightness, contrast, saturation, and horizontal flipping. The test set was kept intact.

The final round was aimed at comparing the best deep model obtained in the previous phase,  ResNet-18, against human experts concerning the task of AMD identification. In this step, twenty real images (ten AMD images and ten non-AMD images) were randomly selected from the test set and provided to human experts for classification purposes. Following that, the same images were submitted to a ResNet-18 for comparison purposes. To allow a fair comparison between humans and deep models, we considered the following measures: standard accuracy (ACC), sensitivity, and specificity.

In this work, we used the StyleGAN2-ADA official source code, and the hyperparameters suggested by Karras et al.~\cite{Karras2020ada}. Concerning StyleGAN2-ADA hyperparameters, we used a batch size of $12$, ADA target equal to $0.8$ (i.e., the probability of using ADA mechanism), the Adam algorithm with a learning rate of $0.0025$, decay values of $0$ and $0.99$, and a convergence error of $1e^{-8}$ for the generator and discriminator. StyleGAN2-ADA framework enables different augmentations (rotation, geometric transformations, and color transformations) and class-conditional training. The output image resolution was set to $256\times256$ pixels. 

\section{Results}
\label{s.experimental_results}

The experimental results are presented in four sub-sections: (i) FID for synthetic image assessment, (ii) human experts detecting synthetic images, (iii) data augmentation assessment, (iv) comparison between human experts and deep models for detecting AMD images, and (v) web-based tool to access and validate the deep learning method model.

\subsection{Synthetic Image Assessment}
\label{ss.image_augmentation}
 
Table~\ref{t.uncgans} presents the FID values for each GAN-based architecture. StyleGAN2-ADA achieved the lowest FID value of $166$, while EBGAN was placed in last and its FID value of $380$ was the highest. The smaller the FID value, the better is the quality of the generated image. Therefore, all further experiments only considered the StyleGAN2-ADA architecture for synthetic image generation. 

\begin{table}[!ht]
    \renewcommand{\arraystretch}{0.90}
    \setlength{\tabcolsep}{9pt}
    \centering
    \caption{Mean FID values for image quality assessment (the best result is highlighted in bold).}
    \label{t.uncgans}
    \begin{tabular}{ccc}
        \toprule
        \textbf{Type} & \textbf{Architecture} & \textbf{FID}
        \\ \midrule
        \multirow{7}{*}{Unconditional} & EBGAN & $380.18$
        \\
        & DCGAN & $326.85$
        \\
        & DRAGAN & $317.82$
        \\
        & WGAN & $307.00$
        \\
        & LSGAN & $305.59$
        \\
        & WGAN-GP & $295.23$
        \\
        & BEGAN & $225.89$
        \\ \midrule
        \multirow{3}{*}{Conditional} & CGAN & $342.59$
        \\
        & ACGAN & $315.36$
        \\
        & \textbf{StyleGAN2-ADA} & $ \textbf{166.17} $
        \\ \bottomrule
    \end{tabular}
\end{table}

\begin{figure}[!ht]
  \centering
  \begin{tabular}{c}
        \includegraphics[scale=0.18]{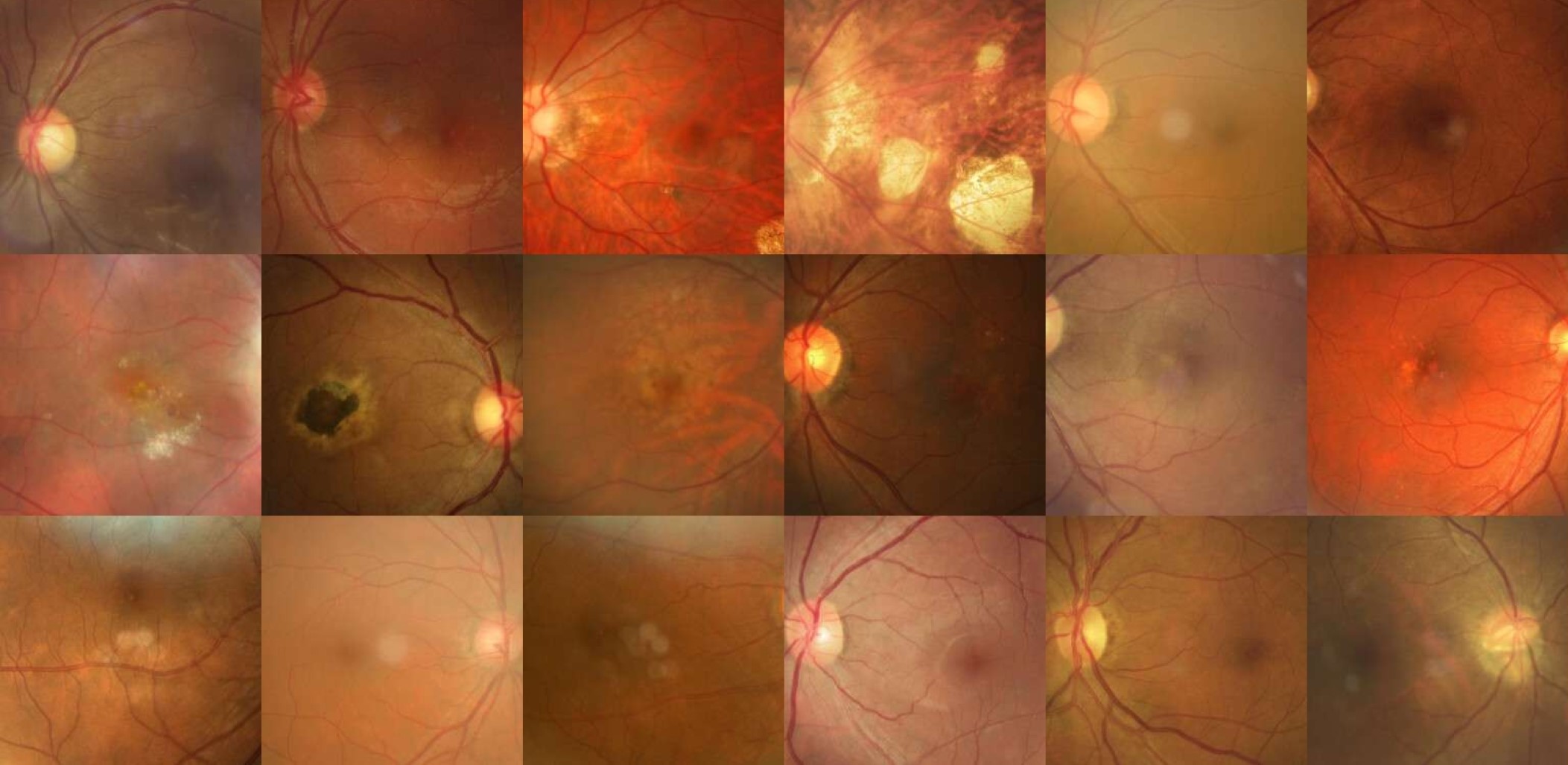} \\
        (a)\\
        \includegraphics[scale=0.18]{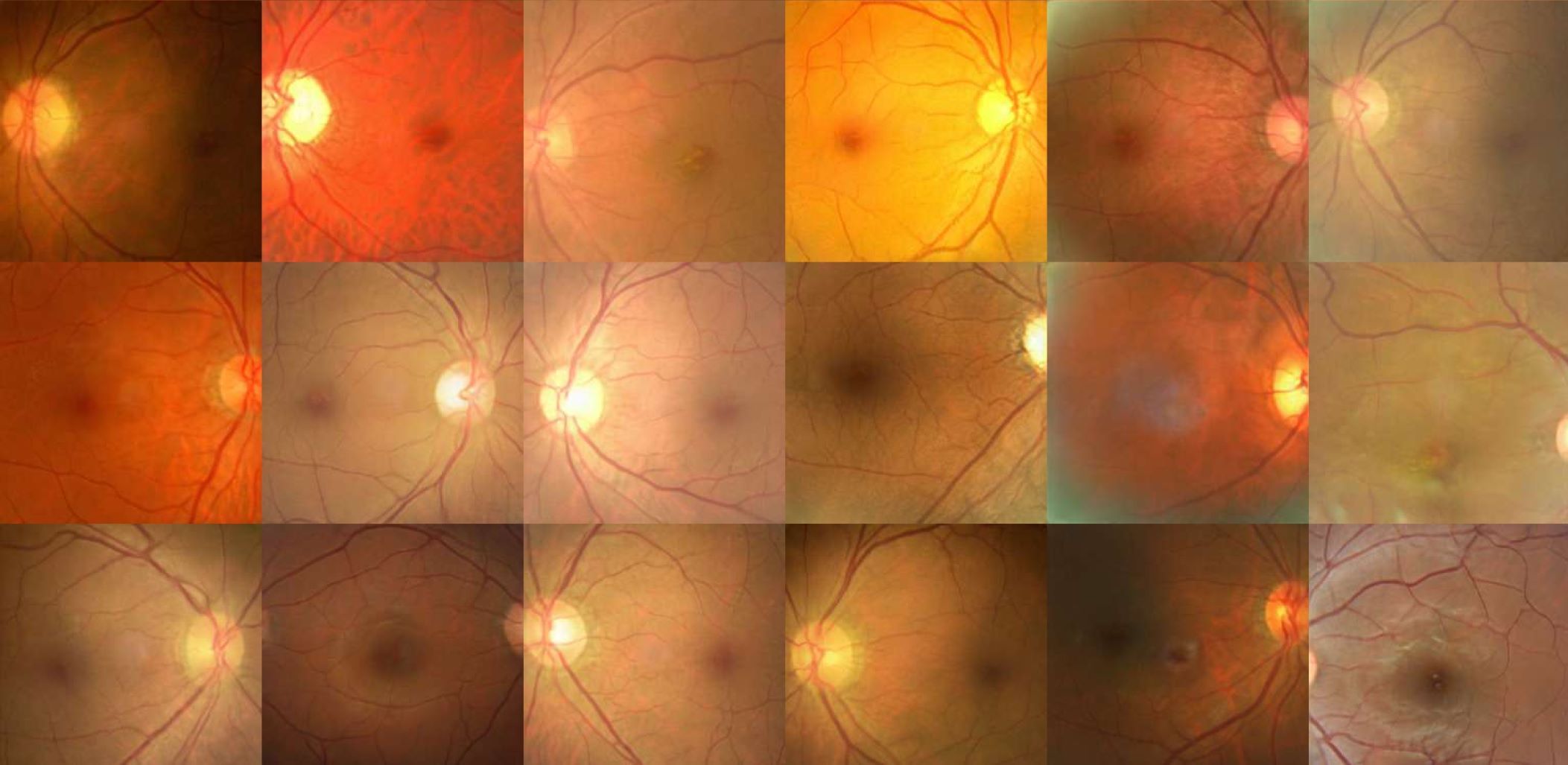} \\ 
        (b)
    \end{tabular}
\caption{  Examples of (a) real retina images extracted from the training dataset, and (b) synthetic images generated by StyleGAN2-ADA. }
\label{f.labels}
\end{figure}

\begin{sloppypar}
Figure~\ref{f.labels} shows a comparison between (a) real images from the training dataset and (b) synthetic images produced by StyleGAN2-ADA. The trained model yields realistic-looking images for both, with and without AMD, conditioned by sampling from latent representations. Visual inspection shows that the generated images are similar to the real images. In the AMD images, macula degeneration is evident.

\end{sloppypar}

\begin{figure}[!ht]
  \centering
  \begin{tabular}{cccc}
      	\includegraphics[scale=0.35]{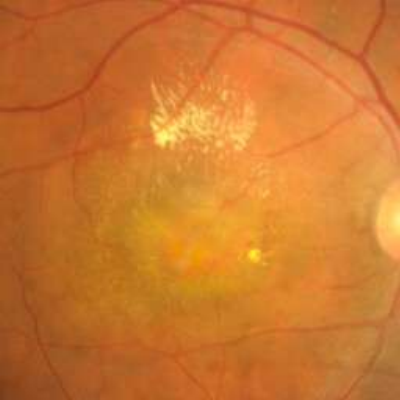} & 
      	\includegraphics[scale=0.35]{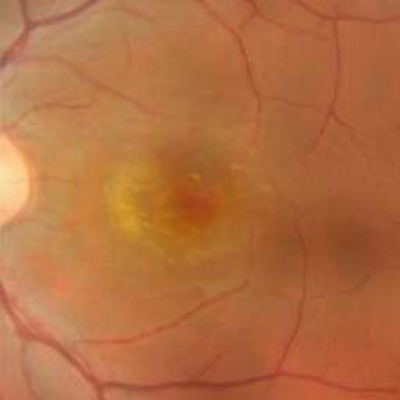} & 
      	\includegraphics[scale=0.35]{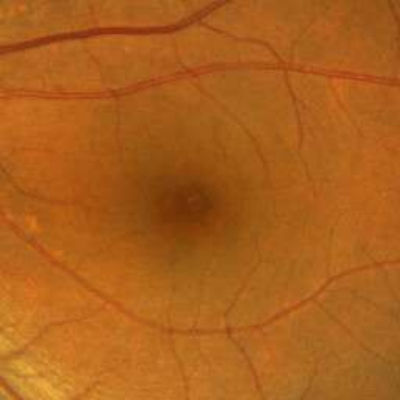} &
      	\includegraphics[scale=0.35]{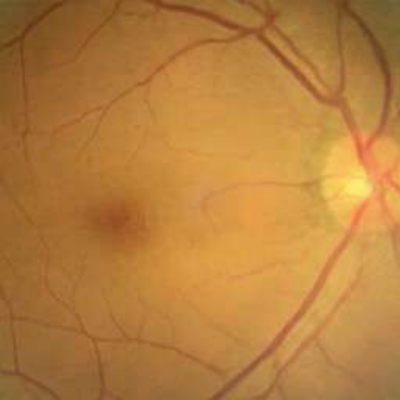} \\ 
      	(a) & (b) & (c) & (d) \\ 
\end{tabular}
\caption{Examples of synthetic and real images for AMD and Non\_AMD. (a) real, positive AMD, (b) synthetic, positive AMD, (c) real, Non-AMD and (d) synthetic, non-AMD.}
\label{f.samples}
\end{figure}

Figure~\ref{f.samples} provides examples of real and synthetic images that are from eyes, positive and negative to AMD. One can observe the high-quality images that were generated for both, AMD and non-AMD images.

\subsection{Distinguishing between Synthetic and Real Images}
\label{ss.synthetic_versus_real}

Table~\ref{t.clinical} presents the outcomes of each clinical expert. For AMD images, the accuracy was $50\%$ (standard deviation of $21.91\%$) for clinician \#1 and $55\%$ (standard deviation of $21.80\%$) for clinician \#2. For Non-AMD images, clinician \#1 achieved an accuracy of $60\%$ (standard deviation of $21.47\%$), and clinician \#2 obtained an accuracy of $50\%$ (standard deviation of $21.47\%$). These results highlight that both clinicians could not differentiate between real and synthetic images.

\begin{table}[!ht]
    \renewcommand{\arraystretch}{0.90}
    \centering
    \scriptsize
    \caption{Synthetic versus real images by humans experts.}
    \label{t.clinical}
    \begin{tabular}{cc|ccccc}
        \toprule &
        & \textbf{ACC} & \textbf{Sensitivity} & \textbf{Specificity} 
        \\ \midrule
        \multirow{2}{*}{AMD} & \textbf{Clinician \#1} & $0.50$ & $0.50$ & $0.50$ 
        \\
        & \textbf{Clinician \#2} & $0.55$ & $0.40$ & $0.70$ 
        \\ \midrule
        \multirow{2}{*}{\shortstack{Non-\\AMD}} & \textbf{Clinician \#1} & $0.60$ & $0.60$ & $0.60$ 
        \\
        & \textbf{Clinician \#2} & $0.50$ & $0.40$ & $0.60$ 
        \\ \bottomrule
    \end{tabular}
\end{table}

\subsection{Data Augmentation Assessment}
\label{ss.image_aug}

\begin{figure}[!ht]
  \centerline{\begin{tabular}{cc}
      	\includegraphics[scale=0.8]{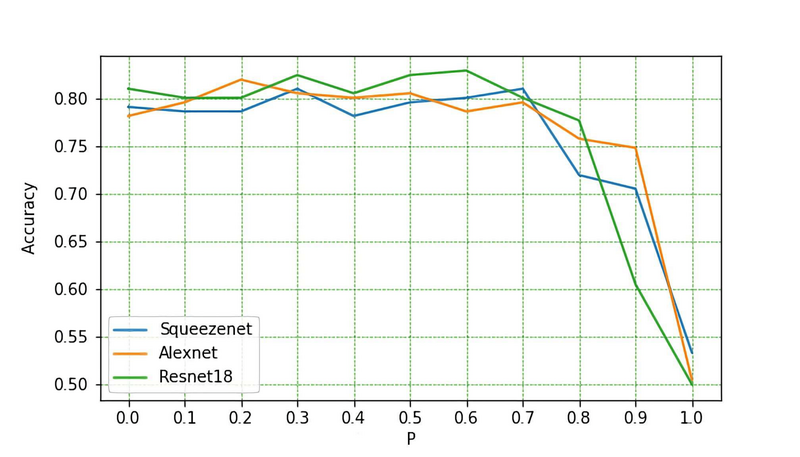} &
      	\includegraphics[scale=0.7]{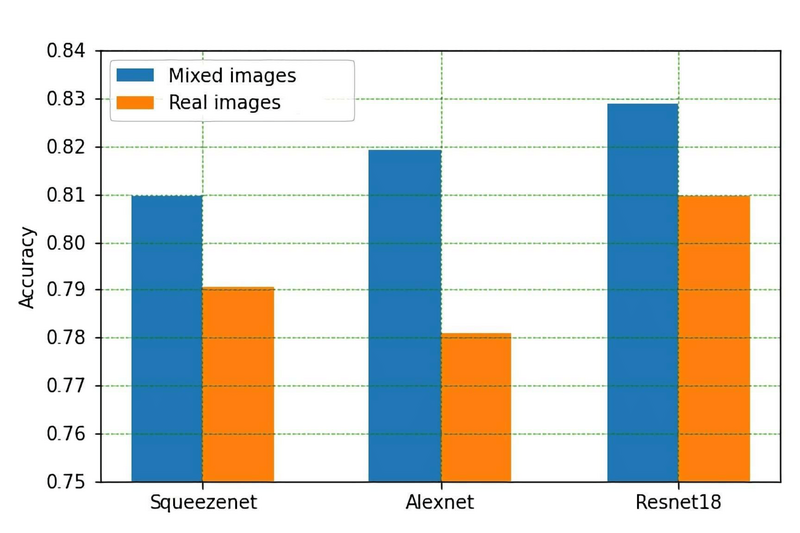} \\ 
      	(a) & (b) \\ 
    \end{tabular}}
    \caption{Accuracy over the test set for different percentages of synthetic image for augmentation purposes (a). Accuracy over the test set concerning ResNet18, AlexNet, and SqueezeNet architectures (b).}
    \label{f.performance}
\end{figure}

In this study we performed on-the-fly data augmentation during training. Figure~\ref{f.performance} shows the accuracy over the test set concerning different percentages ($p$ value) of real images that were replaced by synthetic images. Overall, the accuracy improved when combining synthetic and real images. While the accuracy lies between $50\%$ and $55\%$ when using only synthetic images, and between $78\%$ and $81\%$ when using only real images for training, the combination of both types of images gave the best results. However, this was network dependent: while SqueezeNet accuracy peaked at $81\%$ when using $70\%$ of synthetic images, AlexNet obtained its highest accuracy ($82\%$) when using only $20\%$ of synthetic images. ResNet18 achieved its best of $83\%$ with $60\%$ of synthetic images. In general, the networks performed poorly when the percentage of synthetic images exceeded $70\%$.

Figure~\ref{f.performance}(b) shows the accuracy over the test set using only real ($p=0$) and also mixed data concerning ResNet18, AlexNet, and SqueezeNet architectures. The most significant improvement was by AlexNet,  its accuracy increased by approximately $8\%$, while ResNet18 had highest accuracy (about $83\%$) using a mix of synthetic and real images.


\subsection{Comparison between Human Experts and Deep Models}
\label{ss.comparison_amd_non_amd}

Table~\ref{t.diagnose} presents the comparison of AMD detection by human experts, SqueezeNet, AlexNet and Resnet-18. Overall, the results of both clinicians and deep-learning were similar, with the best performance by deep-learning while the lowest specificity was by clinician \#2. This shows that deep models are at least as good, and may outperform clinicians for diagnosing AMD in eye fundus images.

\begin{table}[!ht]
    \renewcommand{\arraystretch}{1.0}
    \centering
    \scriptsize
    \caption{Comparison between human experts and deep models to classify AMD and real Non-AMD images.}
    \label{t.diagnose}
    \begin{tabular}{c|ccc}
        \toprule & \textbf{ACC} & \textbf{Sensitivity} & \textbf{Specificity}
        \\ \midrule
        \textbf{Clinician \#1} & $0.80$ & $0.80$ & $0.80$ 
        \\
        \textbf{Clinician \#2} & $0.75$ & $1.00$ & $0.50$ 
        \\ 
        \textbf{SqueezeNet} & $0.80$ & $0.80$ & $0.80$
        \\
        \textbf{AlexNet} & $0.80$ & $0.70$ & $0.90$
        \\
        \textbf{Resnet-18} & $0.85$ & $0.90$ & $0.80$
        \\ \bottomrule
    \end{tabular}
\end{table}

\subsection{Web Application}

\begin{figure}[!ht]
    \centering
    \begin{tabular}{cc}
        \includegraphics[width=7.5cm,height=6.5cm,frame]{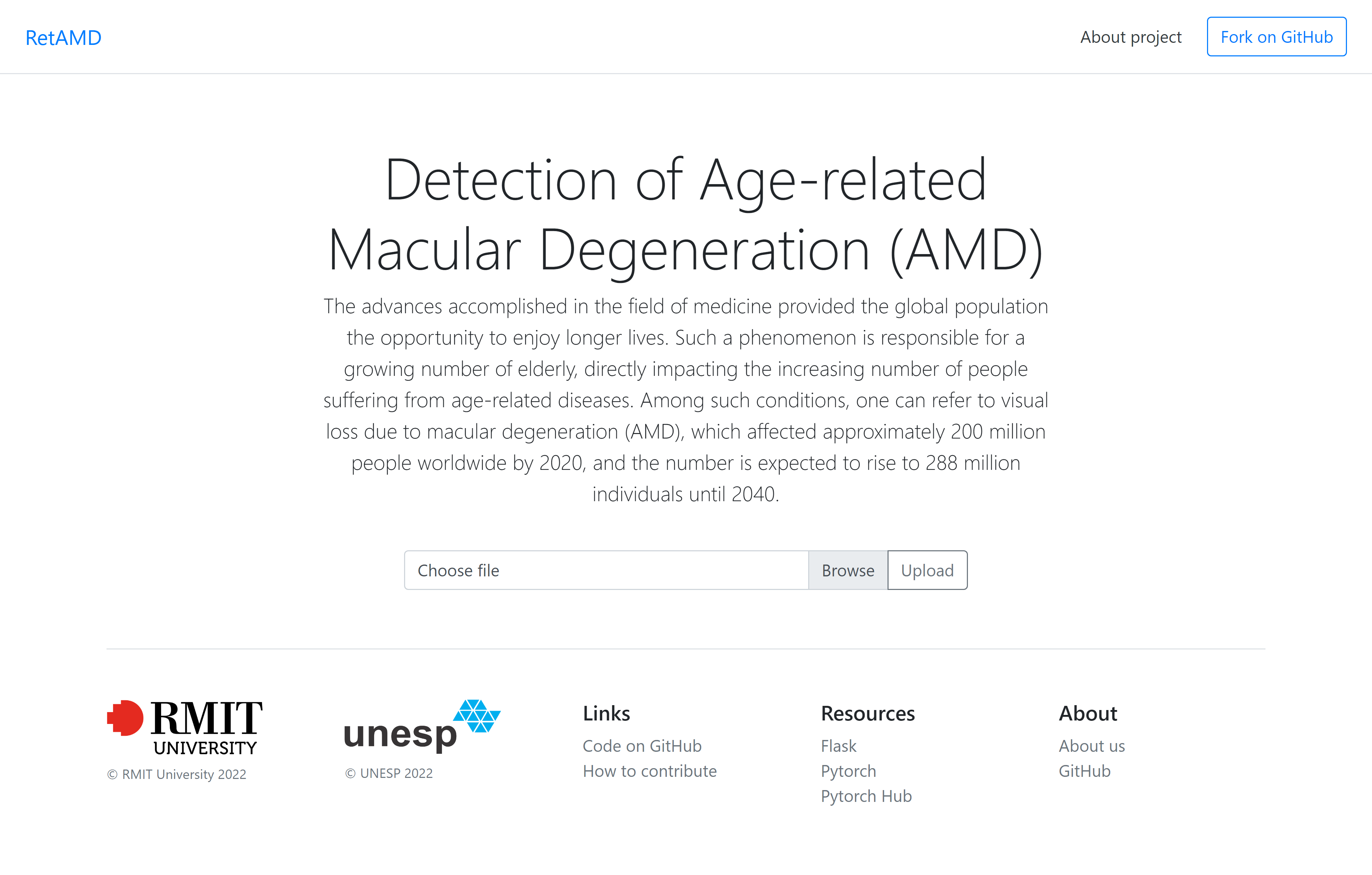}
         &
        \includegraphics[width=7.5cm,height=6.5cm,frame]{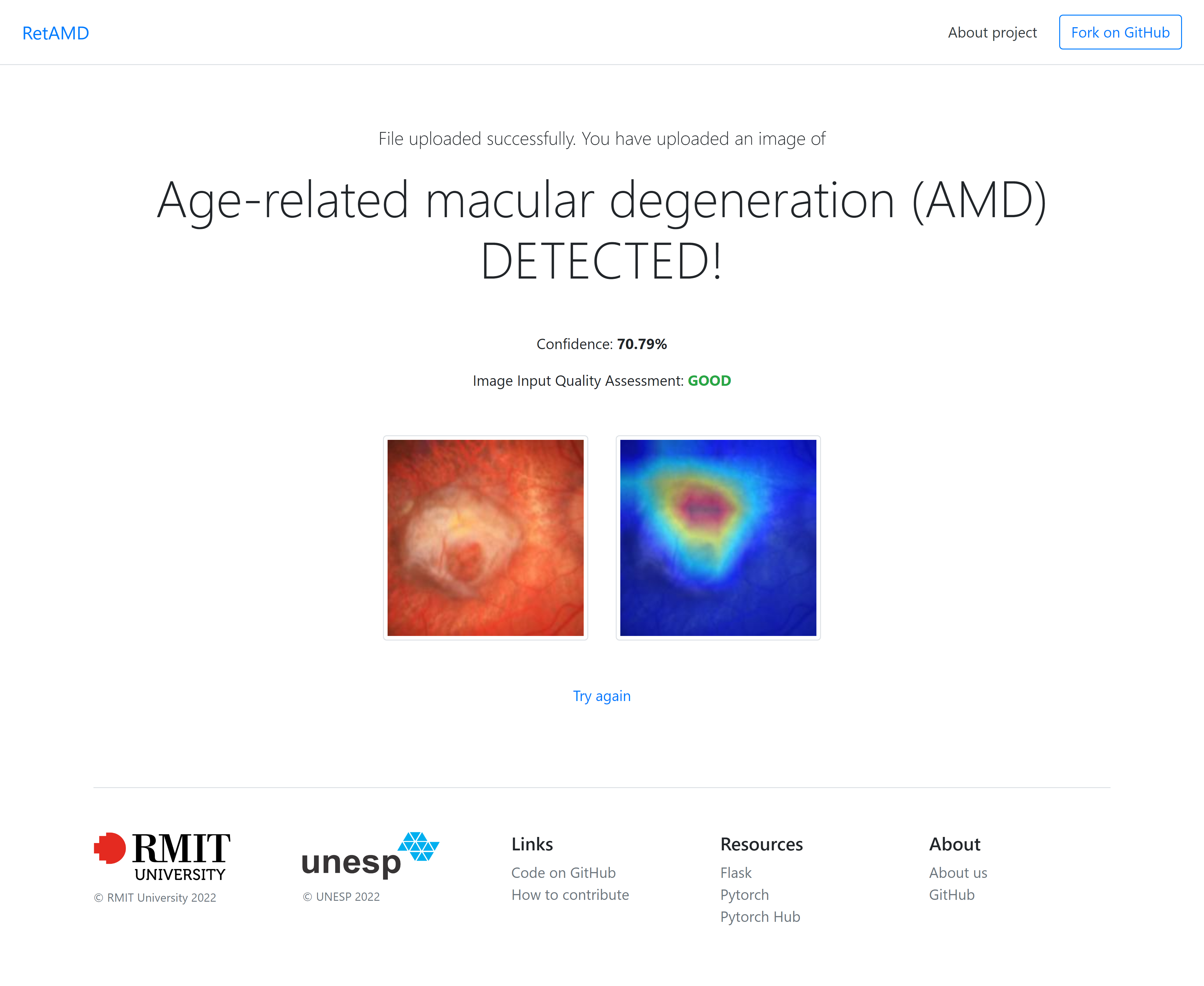}
        \\
         (a) RetAMD home page. 
         &
         (b) Interface showing positive to AMD.
    \end{tabular}
    \caption{Screenshots.}
    \label{f.homepage}
\end{figure}

The web application was designed to be easy to use and accessed by any user. The application's home page is minimalist and features a title along with a brief description of its functionality and a button with which the users can upload an image using the ``browse" button (Figure~\ref{f.homepage}(a)).

Once an image file is selected, the system automatically uploads the image to the cloud, performs the analysis in approximately ten to third seconds (largely based on the internet conditions) displays the output in the form of a figure. The output also informs the user of the success of uploading the file and gives it the label, i.e., AMD detected or not. Additionally, the results also state the level of confidence in the diagnosis and an assessment of the quality of the input image into three quality grades: ``Good", ``Usable" and ``Reject".

One of the main issues regarding deep learning in medicine is the difficulty to interpret the decision mechanisms. On the result screen, users can view two images (Figure~\ref{f.homepage}(b)). The uploaded image is displayed on the left and a heatmap created by GradCAM~\cite{selvaraju2017grad} is displayed on the right. This heatmap highlights the regions that were used in the diagnosis and allows the clinical users to understand the basis of the system's decision. Overall, the results demonstrate the effectiveness and usability of RetAMD in diagnosing AMD.

\subsection{Model Generalization}

The validation of the ResNet18 architecture trained with mixed data ($p=0.6$) are are shown in Table~\ref{t.conf_gridseach} provides the outcomes. The specificity, sensitivity, and accuracy were $80\%$, $85.7\%$, and $82.8\%$, respectively. To test this for generalisability, this deep learning trained model was validated using STARE, a database that was not used during the training phase. 80 images were randomly sampled from STARE; 40 AMD and 40 non-AMD images.

\begin{table}[!ht]
\scriptsize
\begin{minipage}[t]{.5\linewidth}
    \renewcommand{\arraystretch}{0.9}
    \centering
    \begin{tabular}{c|cc}
        \toprule
        & \textbf{AMD} & \textbf{Non-AMD}
        \\ \midrule
        \textbf{AMD} & $90$ & $15$
        \\
        \textbf{Non-AMD} & $21$ & $84$
        \\ \midrule
        \textbf{Specificity} & \multicolumn{2}{c}{$80\%$}
        \\
        \textbf{Sensitivity} & \multicolumn{2}{c}{$85.71\%$}
        \\
        \textbf{Accuracy} & \multicolumn{2}{c}{$82.86\%$}
        \\
        \bottomrule
    \end{tabular}
    \caption{Confusion matrix in our test set.}\label{t.conf_gridseach}
\end{minipage}
\hfill 
\begin{minipage}[t]{.5\linewidth}
    \renewcommand{\arraystretch}{0.9}
    \centering
    \begin{tabular}{c|cc}
        \toprule
        & \textbf{AMD} & \textbf{Non-AMD}
        \\ \midrule
        \textbf{AMD} & $30$ & $10$
        \\
        \textbf{Non-AMD} & $5$ & $35$
        \\ \midrule
        \textbf{Specificity} & \multicolumn{2}{c}{$87.5\%$}
        \\
        \textbf{Sensitivity} & \multicolumn{2}{c}{$75.0\%$}
        \\
        \textbf{Accuracy} & \multicolumn{2}{c}{$81.25\%$}
        \\
        \bottomrule
    \end{tabular}
    \caption{Confusion matrix in STARE.}\label{t.idrid_results}
\end{minipage}
\end{table}

\section{Discussion}
\label{ss.discussion}

Burlina et al have successfully developed a method for ~\cite{burlina2019assessment,burlina2018utility} generating the synthetic images. Their method is based on GAN, and they tested their method by showing that experts were unable to identify the synthetic images. However, their method has been patented and hence not available for being used by others. We have introduced an approach that integrates a deep learning quality assessment model and StyleGAN2-ADA, an extension of the progressive GAN. This approach filters out poor-quality retinal images and generates synthetic medical images that human experts are unable to distinguish from real ones. We have also shown that these images were suitable for increase the performance of deep-learning networks. The results were similar to those reported in the literature using patented technology. While Burlina et al. trained a Progressive GAN over a large number of images positive to AMD, our technique has the potential to generate similar high-quality synthetic images with only a small number of images.

Our method has the potential to enhance the approach outlined by Burlina~\cite{burlina2019assessment,burlina2018utility}. Assessing image quality could lead to the exclusion of poor-quality images. Additionally, implementing on-the-fly augmentation and selecting the appropriate percentage of synthetic images have shown to be fundamental aspects.

ResNet18 architecture trained over real and synthetic images provided the best results, marginally outperforming the human experts' performance. While works of Anh et al.~\cite{ahn2023fundusgan} report significantly better results, these are all from the same database, with similar equipment used for all the images, images having similar quality and people from the similar ethnic group, which however does not represent the real-world situation. Our work has shown the potential of using deep learning over multiple datasets where there are differences in people, equipment and image quality.

One limitation of our study is that it has only handled the binary problem, i.e., AMD versus non-AMD images. However, medical images are often non-binary with more classes. While this study and Burlina et al.~\cite{burlina2019assessment} utilized the expertise of two clinical professionals for image evaluation, it's important to acknowledge that having more experts such as three experts can make the study stronger and tests for the generalisability of the model. Further, three experts permits the disagreements being resolved by voting which tests the strength of the study. 

Another limitation of this study is that it used a single-source dataset with 80 images for validation purposes. However, this may be not sufficient to test the model's generalisability.

Different studies have presented results reaching more than 90\% accuracy~\cite{pead2019automated, leng2023deep}. However, the aim of this study was not to get the best results but for the model to be not limited to any one database. Thus, the focus was on training on a mixed dataset to increase generalization and test this using a different dataset. Another objective was also to develop a web-based tool that provides a model capable of identifying AMD in a wide variety of images from different sources. Our model works on standard CPU computers, which makes it suitable for inexpensive deployment. One limitation in our work is that we have used traditional architectures and not used the number of layers reported nor some recent advances in the state-of-the-art techniques such as used by Anh et al.~\cite{ahn2023fundusgan}, Grassmann et al.~\cite{grassmann2018deep}, Govindaiah et al.~\cite{govindaiah2018new}, Keel et al.~\cite{keel2019development}, Bhuiyan et al.~\cite{bhuiyan2020artificial}. This decision was influenced by a common drawback of state-of-the-art convolutional neural networks, which is their complexity. They were typically developed to be trained in extensive datasets, resulting in overfitting when applied to small datasets. However, this study focused on developing a system suitable for being used for web-based applications.

We have made available an online system with this trained network so that anyone can use it and test it, simply by uploading images. The software automatically labels the images as positive or negative to AMD. We have also provided the source code of the entire software and it is available publicly to facilitate researchers to use this as it is, or improve it. We are focused on fostering partnerships to facilitate and conduct research towards the usage of deep-learning to generate and recognize medical images. 

The potential of diffusion models~\cite{croitoru2023diffusion}, known for their advanced capabilities in generating high-quality and diverse images, presents exciting future research in AMD and other ophthalmology diagnoses. These models should be considered for future development.

\section{Conclusion and Future Works}
\label{s.conclusion} 

We have employed a retinal image quality assessment model in preprocessing step. We have compared a number of synthetic medical image generation techniques and found StyleGAN2-ADA to be the most suitable using which we have developed a method to generate synthetic images. We have investigated the use of the synthetic images obtained using examples from publicly available databases to train the model for distinguishing between AMD and healthy eyes. We found that experienced clinical experts were unable to differentiate between synthetic and real images. We have tested the model for generalisability by training the model using images from three databases and validated it using a fourth database.  We also have demonstrated that the classification accuracy of deep learning networks marginally outperformed clinical experts in separating the AMD and Non-AMD retinal images.

We have made the source code for generating the synthetic images publicly available to facilitate joint research in the field. We have also provided free access through this paper for the online use of the AMD detection model. This will facilitate future work to broaden the scope for detecting the severity of AMD, and for differentiating from other diseases. For generating synthetic medical images, there is the need to consider a broader range of deep architectures and the effectiveness of heatmaps helping the clinicians.

\section{Data availability}
\label{s.data_availability}

\begin{sloppypar}
The iChallenge-AMD dataset can be found in~\url{https://ai.baidu.com/broad/introduction?dataset=amd}, while ODIR-2019 dataset is available on~\url{https://odir2019.grand-challenge.org/dataset/} and \textbf{RIADD} is available on ~\url{https://riadd.grand-challenge.org/Home/}.
\end{sloppypar}

\section{Code availability}
\label{s.code_availability}

\begin{sloppypar}
The official code for StyleGAN2-ADA is available at~\url{https://github.com/NVlabs/stylegan2-ada-pytorch}. Our implementation of Deep Convolutional GAN (DCGAN), Least Squares Generative Adversarial Networks (LSGAN), Wasserstein GAN (WGAN), Wasserstein GAN with Gradient Penalty (WGAN-GP), Deep Regret Analytic Generative Adversarial Networks (DRAGAN), Energy-based Generative Adversarial Network (EBGAN), Boundary Equilibrium Generative Adversarial Networks (BEGAN), Conditional GAN (CGAN), and Auxiliary classifier GAN (ACGAN) are available for download at~\url{https://github.com/GuiCamargoX/gans-pytorch}. All source code concerning image processing, Style-GAN2-ADA data generation, and the pre-trained networks are available at \url{https://github.com/GuiCamargoX/synthetic-retina-amd}.  The web application can be find in \url{https://amdfundus.space/}. The official Flask code of is available for download in \url{https://github.com/GuiCamargoX/RetAMD}.
\end{sloppypar}

\section*{Author Information}

\subsection{Contributions}

G.O. developed the algorithms and performed the analysis. 
J.P. and D.K. conceptualized and designed the project. 
G.R., L.P., D.P. managed the data and designed the software.
H.K. was responsible for the clinical assessment of the images and script. 
All authors contributed to writing the manuscript.
D.K, G.O. and JP finalized the manuscript.

\section*{Conflict of Interest declarations}
The authors declare no conflict of interest for this manuscript.

\section*{Acknowledgment}
The authors thank Dr. Ione F. Alexim from BP Hospital - \textit{A Beneficência Portuguesa de São Paulo}, who provided her eye fundus grading expertise to the project. We are grateful to S\~ao Paulo Research Foundation (FAPESP) under the grants \#2013/07375-0, \#2014/12236-1, \#2018/15597-6, \#2019/00585-5, \#2019/07665-4, \#2019/02205-5, and \#2023/10823-6, to the Brazilian National Council for Research and Development (CNPq) grants \#307066/2017-7, \#427968/2018-6, \#309439/2020-5, and \#88887.606573/2021-00, as well as the Engineering and Physical Sciences Research Council (EPSRC) grant EP/T021063/1 and its principal investigator Ahsan Adeel. We also acknowledge the funding from Promobilia Foundation, Sweden, and SPARC grant (P-134, 2019), Department of Biotechnology (India) for financial their support.



\bibliographystyle{unsrt}

\bibliography{references}

\end{document}